\documentclass[twocolumn, prd, 10pt, floatfix]{revtex4-1}

\usepackage{amsmath}
\usepackage{amssymb}
\usepackage{mathtools}
\usepackage{bm}
\usepackage{graphicx,color}
\usepackage{hyperref}
\usepackage{braket}
\usepackage{textcomp}
\usepackage{tikz}
\usepackage{pstricks}
\usepackage{ragged2e}
\usepackage[utf8]{inputenc}
\usepackage{wrapfig}
\usepackage{booktabs}
\usepackage[english]{babel}
\usepackage{epstopdf}

\mathtoolsset{showonlyrefs}

\hypersetup{
  colorlinks   = true, 
  urlcolor     = blue, 
  linkcolor    = blue, 
  citecolor   = red, 
}

\newcommand{\tr}{\operatorname{tr}}

\newcommand{\ic}{\ensuremath{\mathrm{i}}}

\renewcommand{\vec}[1]{\ensuremath{\bm{#1}}}

\begin{document}

\title{Matrix product states and the nonabelian rotor model}


\author{Ashley Milsted}
\email[]{ashley.milsted@itp.uni-hannover.de}
\affiliation{Leibniz Universit\"at Hannover, Institute of Theoretical Physics, Appelstrasse 2, D-30167 Hannover, Germany}

\date{\today}

\begin{abstract}
We use uniform matrix product states (MPS) to study 
the (1+1)D $O(2)$ and $O(4)$ rotor models, which are equivalent
to the Kogut-Susskind formulation of matter-free nonabelian lattice gauge theory 
on a ``hawaiian earring'' graph for $U(1)$ and $SU(2)$, respectively.
Applying tangent space methods
to obtain ground states and determine the mass gap and the $\beta$ function, 
we find excellent agreement with known results, locating the
BKT transition for $O(2)$ and successfully entering the asymptotic weak-coupling
regime for $O(4)$.
To obtain a finite local Hilbert space, we truncate in the space of generalized Fourier
modes of the gauge group, comparing the effects of different cutoff values.
We find that higher modes become important in the crossover and 
weak-coupling regimes of the nonabelian theory, where entanglement also
suddenly increases. This could have important consequences for TNS
studies of Yang-Mills on higher dimensional graphs.
\end{abstract}

\maketitle

\section{Introduction}

Nonabelian gauge theories describe the interactions responsible
for most of the matter we experience in our everyday lives.
In particular, they explain the hadrons |
bound states of quarks | which include neutrons and protons among
their most famous examples \cite{peskin_1995}. Curiously, the quarks inside
hadrons behave as free particles for the purposes of high-energy
scattering (asymptotic freedom), yet they are never observed in isolation
(confinement \cite{greensite_2011}).
These properties are also present in matter-free nonabelian gauge
theory (pure Yang-Mills theory). 
This apparently simple theory, despite its huge symmetry group of local
gauge transformations, resists exact solution and must so far be
approached with approximate methods such as perturbation theory and numerical
tools such as Monte Carlo sampling \cite{creutz_1985}, albeit with convincing successes,
such as the determination of Hadron masses using lattice simulations \cite{durr_2008}.

Monte Carlo techniques are also extremely useful in condensed matter physics and
advances have benefited both fields. However, in recent decades new, highly general
techniques have arisen in condensed matter and quantum information that open up 
whole new avenues of numerical investigation. These techniques exploit 
Tensor Network States (TNS) \cite{bridgeman_2016}, which efficiently represent many-body states 
with limited \emph{entanglement}. The best-known example is the
density-matrix renormalization group (DMRG) \cite{white_1992}, 
which can be viewed \cite{schollwock_2011} as a variational
algorithm applied to one-dimensional TNS, also known as Matrix Product States (MPS)
\cite{fannes_1992, *rommer_1997, *vidal_2004}.
These methods are inherently free of the sign problem that plagues Monte Carlo
sampling \cite{vonderlinden_1992} and offer themselves naturally to simulation of real-time dynamics.

In the last years, as TNS techniques have advanced (higher dimensions,
more sophisticated networks, improved numerical tools)
\cite{verstraete_2008, *vidal_2009, *haegeman_2013a, *orus_2014},
efforts have increased to transfer their successes in condensed matter
to quantum field theory, particularly with an eye toward nonabelian
gauge theory. Important steps in this direction include ground state,
real-time, and finite-temperature simulations of $\phi^4$ theory
\cite{sugihara_2004, milsted_2013a}, the Schwinger model
\cite{byrnes_2002, banuls_2013, *banuls_2013a, *banuls_2016,
  rico_2014, buyens_2014, *buyens_2015}, $SU(2)$ gauge theory with
matter \cite{kuhn_2015} in (1+1)D, and quasi-one-dimensional abelian
gauge theories \cite{sugihara_2005, *sugihara_2005a}, all using MPS or
DMRG, as well as proposals for representing lattice gauge theory
states in higher dimensions, with a view toward numerics as well as
analytics \cite{tagliacozzo_2011, silvi_2014, tagliacozzo_2014,
  haegeman_2015, milsted_2015}.  The tensor renormalization group
(TRG) algorithm \cite{levin_2007, *gu_2010, *gu_2013} has also been
applied to $\phi^4$ theory \cite{shimizu_2012}, the Schwinger model
\cite{shimizu_2014}, and the $O(2)$ and $O(3)$ models \cite{yu_2014,
  unmuth-yockey_2014}.

\begin{figure}
  \includegraphics{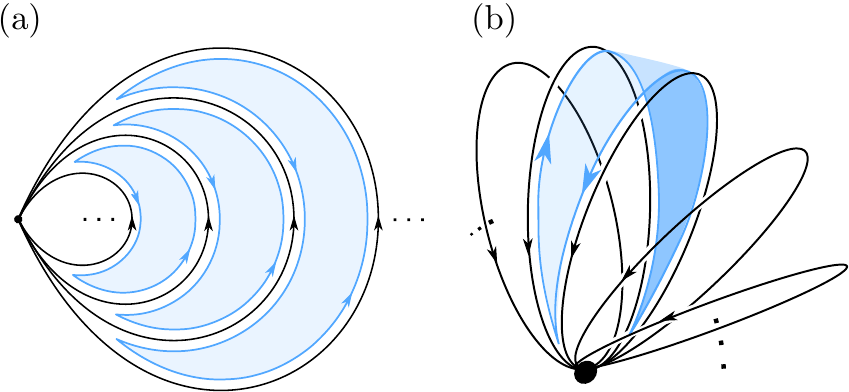}
  \caption{\label{fig:petal_graph} Illustration of lattice gauge theory (a) on a 
           ``hawaiian earring'' and (b) the same theory visualized differently 
           as living on the surface of a 3D object.
           The Hamiltonian is the Kogut-Susskind \cite{kogut_1975} 
           formulation of lattice gauge theory for a gauge group $G$ | 
           example plaquette operators are shown in blue.
           The Hilbert space $\mathcal{H}$ (including nonphysical states) 
           is made up of systems living on the (black) edges
           $\mathcal{H}_\text{edge} \cong L^2(G)$.
           For the gauge groups $G \cong U(1)$ and $G \cong SU(2)$ this model is equivalent to the
           (1+1)-dimensional quantum rotor model \cite{sachdev_2011} 
           for the rotation groups $O(2)$ and $O(4)$, respectively.
           }
\end{figure}

\subsection{This work}
Here we present an MPS study of the $O(2)$ and $O(4)$ quantum rotor models 
in $1+1$D,
which are equivalent, respectively, to the $U(1)$ and $SU(2)$ principal chiral field (PCF) models. The PCF model is in turn equivalent to a pure lattice gauge theory on a ``hawaiian earring'' graph in the Kogut-Susskind formulation 
(see Figure \ref{fig:petal_graph}), insofar as we do not restrict to gauge-invariant states.

The Hamiltonians possess a global gauge-group symmetry
rather than a local gauge symmetry, but nevertheless have a lot in common with
Yang-Mills on more sophisticated graphs. Most importantly, the $O(N > 2)$ models are
known to possess a single, gapped phase ending at the weak-coupling limit 
$g \rightarrow 0$ \cite{sachdev_2011}. This is also observed in simulations of (3~+~1)D 
nonabelian lattice gauge theory, in which the gapped phase is confining \cite{greensite_2011}.
In contrast, the $O(2)$ model has a phase transition (of Berezinskii-Kosterlitz-Thouless (BKT)
type \cite{kosterlitz_1973}) at finite coupling, transitioning into a 
deconfined, gapless phase at weaker couplings.

The continuum limit of the rotor models, the so-called $O(N)$ nonlinear sigma model
\cite{sachdev_2011}, can be solved using the Bethe ansatz for $N > 2$ making the lattice 
weak-coupling scaling of the mass gap computable \cite{hasenfratz_1990}. 
The $O(N)$ model has also been thoroughly investigated using
strong-coupling expansions \cite{hamer_1979, hornby_1985}, which operate on the same
$1+1$D Hamiltonian model we study here, as well as 
high-temperature expansions (for example \cite{butera_1993, *butera_1996}) and Monte Carlo numerics 
(\cite{fox_1982, wang_1993, weisz_2001, alet_2003} is an incomplete selection) applied to the 
2D classical $O(N)$ model. Lanczos diagonalization with finite-size-scaling has also been used \cite{roomany_1980}.
In this work, we use uniform MPS to represent infinite, translation invariant states,
applying the nonlinear conjugate gradient method \cite{milsted_2013a} 
to obtain ground states and the MPS tangent space as an ansatz for 
low energy excitations \cite{haegeman_2012, haegeman_2013}, 
determining the mass gap and the $\beta$ function
at finite couplings and thus obtaining the phase diagram.

\section{The model}

\subsection{Kogut-Susskind Hamiltonian}
The Kogut-Susskind Hamiltonian \cite{kogut_1975} on the ``hawaiian earring'' graph
is given by
\begin{equation} \label{eq:H_KS}
	H_{\text{KS}}(g) = \frac{\sqrt{\eta}g^2}{2 a} \sum_{k=-\infty}^{+\infty} E^2_k - \frac{2\sqrt{\eta}}{g^2 a}\sum_{k=-\infty}^{+\infty} \text{Re}(\tr (u_k u_{k+1}^\dagger)),
\end{equation}
where $g$ is the coupling, $a$ the lattice spacing, and $\eta$ an anisotropy parameter
required to ensure the renormalized theory is Lorentz-invariant in the continuum limit
\cite{shigemitsu_1981}.
The Hilbert space is the tensor product of spaces 
\begin{equation}
  \mathcal{H}_k = L^2(G)
\end{equation}
assigned to each edge $e$ in the graph and $G$ is the gauge group.
We define $\lambda_\alpha$ to be the
Hermitian generators of $G$ (for $SU(2)$ these are the Pauli matrices
$\lambda_\alpha = \frac{1}{2} \sigma_\alpha$ with $\alpha=1,2,3$, 
for $U(1)$ there is only one $\lambda = 1$). 
The operator 
\begin{equation}
  E_k^2 = \sum_\alpha E_{\alpha,k}^2
\end{equation}
is the quadratic Casimir operator
representing the kinetic energy within the gauge group at edge $k$. The $E_\alpha$ represent the infinitesimal group action 
\begin{equation}
E_\alpha := \partial_\epsilon \left. L_{e^{\ic \epsilon \lambda_\alpha}} \right|_{\epsilon=0},
\end{equation}
where $L_x$ implements rotations from the left, acting on a ``position'' basis as 
\begin{equation}
L_x |v\rangle = |xv\rangle
\end{equation}
for $x,v \in G$.
The $u_{ij}$ are gauge group position operators defined as 
\begin{equation}
  u_{ij}|v\rangle = t(v)_{ij}|v\rangle,
\end{equation}
with $t(v)$ an irrep of $G$ (we choose $e^{\ic \theta}$ for $U(1)$ and the spin-half
representation for $SU(2)$).
This results in the commutator
\begin{equation}
\left[E_\alpha, u_{ij} \right] = \sum_{j'} \lambda_{\alpha, ij'} u_{j'j}.
\end{equation}

\subsection{Quantum Rotor Hamiltonian}

The model \eqref{eq:H_KS} is known to be equivalent to a chain of coupled $O(N)$ rotors 
(see e.g.~\cite{shigemitsu_1981}), given by
\begin{equation} \label{eq:H_R}
 H_{\text{R}}(\tilde g) = \frac{\sqrt{\eta} \tilde g}{2 a} \sum_{k=-\infty}^{+\infty} \vec{J}^2_k - \frac{\sqrt{\eta}}{\tilde g a}\sum_{k=-\infty}^{+\infty} \vec{n}_k \cdot \vec{n}_{k+1},
\end{equation}
where $\vec{n}_k$ is a $N$-dimensional unit vector representing the $k$th rotor and $\vec{J}_k^2$ is the 
rotor kinetic energy. The normalization of $\vec{J}^2$ is chosen to match \cite{hamer_1979}. 
The Hamiltonian~\eqref{eq:H_R} is manifestly invariant under a global $O(N)$ symmetry.
The relations which underlie $H_{\text{KS}} = H_{\text{R}}$ are given in Table~\ref{tab:rotor-vs-ks}.

\begin{table}[h]
    \centering
    \bgroup
    \def\arraystretch{1.2}
    \begin{tabular}{|l|l|l|}
    \hline
    \bf $H_{\text{R}}, O(N)$ & \bf $H_{\text{KS}}, G \cong U(1)$ & \bf $H_{\text{KS}}, G \cong SU(2)$ \\
    \hline
    $N$ & $2$ & $4$ \\
    $\tilde g$ & $g^2 / \sqrt{2}$ & $g^2 / 4$ \\
    $\vec{J}^2$ & $E^2$ & $4 E^2$ \\
    $\vec{n}_\mu$ & $(\mathrm{Re}(u), \mathrm{Im}(u))_\mu$ 
     & $-\ic\tr({\lambda_\mu} u)$ \\
    & $\mu=1,2$ & $\mu = 0 \dots 3,\; \lambda_0 = \frac{\ic}{2} \mathbb{I}$ \\
    \hline
    \end{tabular}
    \egroup
    \caption{This table shows how quantities in the rotor model~\protect\eqref{eq:H_R} 
             must be set to obtain equivalence to the Kogut-Susskind 
             model~\protect\eqref{eq:H_KS} for the gauge groups $G \cong U(1)$ and $G \cong SU(2)$.
             }
    \label{tab:rotor-vs-ks}
\end{table}

The continuous position basis $|v\rangle$ does not lend itself to use with MPS numerics,
for which we require a discrete, finite basis $\cong \mathbb{C}^d$.
Instead we make use of the generalized Fourier basis given by the Peter-Weyl theorem \cite{hall_2015},
in which the matrix elements of the irreducible representations (irreps) of a 
compact Lie group $G$ label generalized Fourier modes $|ij\rangle_l$ (the matrix 
element $i,j$ of irrep $l$).
The kinetic term $E^2$ is diagonal in this basis, 
with 
\begin{equation}
  E^2|ij\rangle_l = l(l+1)|ij\rangle_l, \quad l \in \frac{1}{2} \mathbb{Z}^*
\end{equation} 
for $SU(2)$ and 
\begin{equation}
  E^2|n\rangle = n^2|n\rangle, \quad n \in \mathbb{Z}
\end{equation}
for $U(1)$.
In the strong coupling regime $g^2 \gg 1$, the $E^2$ term strongly penalizes
higher irreps, so we can neglect them to good approximation at larger $g^2$, 
expecting them to become more relevant as we near weak coupling.
Importantly, truncating the basis at a certain irrep level (Fourier mode)
does not prevent representation of states invariant under the global gauge-group
symmetry (since rotations do not mix irreps). 

\section{Numerical methods}

The uniform MPS variational class consists of states
\begin{equation} \label{eq:uMPS}
  |\Psi(A)\rangle = \sum_{\vec{s}=1}^d v_L^\dagger \left[\prod_{k=-M}^{+M} A^{s_k} \right] v_R |s_{-M} \dots s_0 \dots s_{+M} \rangle,
\end{equation}
where $A^s$ is a $D \times D$ matrix and $d$ the dimension of the chosen local basis and we make the chain length infinite: $M \rightarrow \infty$. For a given basis element $\vec{s} = (s_{-M},\dots,s_{+M})$, the matrix product in square brackets determines the value of $\langle s_{-M} \dots s_0 \dots s_{+M} |\Psi(A)\rangle$, hence the name ``matrix product states.'' Two key features of MPS are the efficiency with which quantities such as expectation values of local operators and correlation functions can be computed (requiring $\mathcal{O}(D^3)$ multiplications), and the fact that the restriction to MPS form serves only to limit the amount of entanglement that can be present in the state. The dimension $D$ is called the bond dimension and serves to control the degree of spatial correlations, placing an upper bound $S \le \log D$ on the entanglement entropy, for example. For more background, see for example \cite{bridgeman_2016}.

To make the limit $M \rightarrow \infty$ behave appropriately, we require $A$ such that the \emph{transfer 
operator} 
\begin{equation}
  E := \sum_s A^s \otimes \overline{A^s}
\end{equation}
has spectral radius $\rho(E) = 1$ with a unique eigenvalue of largest magnitude
(injectivity) equal to one. 
With this condition, the boundary vectors $v_L$ and $v_R$ drop out of
all relevant calculations and $\langle \Psi(A) | \Psi(A) \rangle = 1$.
For more details on using infinite, uniform MPS, see \cite{haegeman_2013a}.

We set $d$
to accommodate the dimensions of all Fourier modes up to a cutoff.
With $U(1)$ all irreps are one-dimensional and we may label Fourier modes as $n \in \mathbb{Z}$, 
so that a cutoff is given by $|n| \le n_{\max}$ and 
\begin{equation}
  d = 2n_{\max} + 1.
\end{equation} 
For $SU(2)$ we must set 
\begin{equation}
  d = \sum_{l = 0}^{l_{\max}} \dim(V_l)^2 = \sum_{l = 0}^{l_{\max}} (2l + 1)^2,
\end{equation}
with $l=0,\frac{1}{2},1,\frac{3}{2}, \dots \;$.

In this study, we use values of $n_{\max}$ up to $10$ and $l_{\max}$ up to $2$.
The former requires $d=21$, while the latter implies $d=55$, 
which is unusually high for MPS numerics.
In the algorithms of \cite{haegeman_2011, haegeman_2012, milsted_2013a}
the cost of computations involving nearest-neighbor operators,
such as the potential term $u_e u_{e+1}^\dagger$, scales as $\mathcal{O}(d^4)$. 
We reduce this to $\mathcal{O}(d^2 m)$, where
$m$ is the number of terms in the tensor product decomposition, by implementing
them as two-site matrix product operators \cite{schollwock_2011} of dimension $m$,
where $m \le 4$ for our purposes. 
We further accelerate our implementation by computing the iterands of iterative
parts using general purpose graphics processing units (GPGPU's)
\footnote{In particular, we use the operations provided by the CUBLAS library
on NVIDIA Tesla K20 devices.}. 

With these optimizations, we apply the nonlinear conjugate gradient (CG) method
to obtain ground states \cite{milsted_2013a}, 
with the time-dependent variational principle \cite{haegeman_2011}
used in a pre-optimization step to provide good starting points for the CG algorithm. 
We converge all states up to $||P_{T(A)}H|\Psi(A)\rangle|| \le 10^{-8}$,
where $P_{T(A)}$ projects the energy gradient vector onto the MPS tangent
space at $|\Psi(A)\rangle$.
We then obtain low-lying excited states using the methods of \cite{haegeman_2012},
always operating directly in the space of infinite, uniform MPS.
All MPS algorithms used here are implemented as part
of the open source \emph{evoMPS} project \cite{milsted_2015a}.

\section{Results}

\subsection{Phases observed}

Since our choice of truncated basis is most appropriate at strong coupling,
we study the system starting at $1/\tilde g \rightarrow 0$ and then approach
weak-coupling as far as possible, whilst maintaining accuracy. We find, for
both the $O(2)$ and the $O(4)$ rotor, that the MPS approximate ground state 
breaks the global $O(N)$ symmetry for $1/\tilde g < 1/\tilde g_{\text{SB}}$ for constant, 
finite $D$. The symmetry-breaking, for the values of $D$ in use, is confined
to a relatively narrow region of parameter space.
Since the breaking of a continuous symmetry is forbidden by the Mermin-Wagner theorem
\cite{mermin_1966, *coleman_1973}, this must be a symptom of finite-entanglement effects
\cite{wang_2011, *draxler_2013, *zauner_2015}:  
The bond dimension needed to accurately represent the symmetric state must suddenly grow
as we approach weak coupling.

\begin{figure}[h]
  \includegraphics[width=1\linewidth]{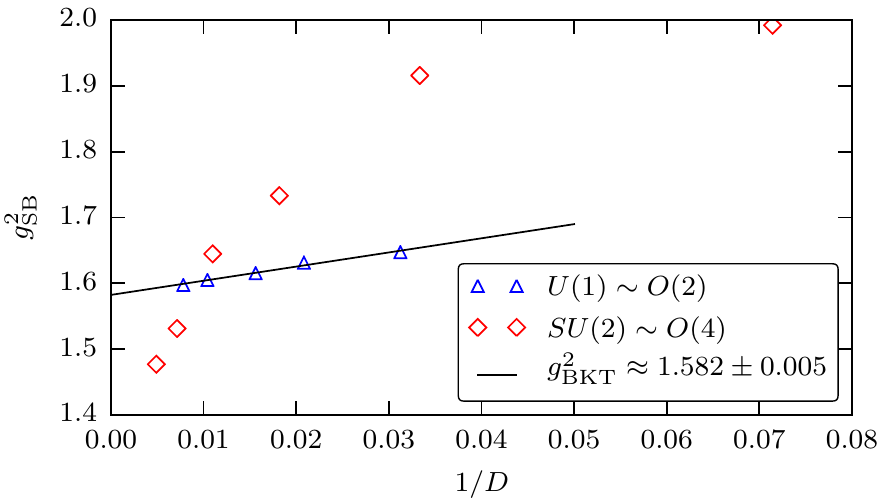}
  \caption{\label{fig:SB_extrap} The location $g^2_{\text{SB}}$ of MPS ground state
           symmetry breaking (SB) as a function of the bond dimension $D$.
		   We plot the Kogut-Susskind coupling $g^2_{\text{SB}}$ rather than 
		   $\tilde g_{\text{SB}}$ to aid comparison.
           Values of $g^2_{\text{SB}}$ were found using bisection up to
           a precision in $g^{-2}$ of $\pm 0.005$.
           For $U(1) \sim O(2)$, $g^{2}_{\text{SB}}$ tends toward a finite value. This value should correspond to the location of the BKT transition \cite{sachdev_2011}. 
		   In terms of the $O(2)$ rotor parameter the fit gives 
		   $\tilde g_{\text{BKT}}= 1.119 \pm 0.004$.
           For $SU(2) \sim O(4)$, the transition does not converge for the data available.
           This is consistent with it occurring at $g = \tilde g = 0$.
           }
\end{figure}

The $O(2)$ rotor is known to possess a gapless phase at weak coupling,
characterized by algebraically decaying correlations, such that the
correlation length is infinite \cite{sachdev_2011}. As defined above, a uniform MPS would
require $D \rightarrow \infty$ to accurately represent such a ground state,
thus explaining nonphysical symmetry-breaking at finite $D$.
The existence of a phase transition at finite $\tilde g$ also explains
the narrowness of the region where symmetry-breaking begins.
We expect the symmetry-breaking location $\tilde g_{\text{SB}}(D)$ to converge to 
the location of the phase transition as $D \rightarrow \infty$
and indeed this convergence can be seen in Figure \ref{fig:SB_extrap},
where the extrapolated transition point $\tilde g = 1.119 \pm 0.004$ 
agrees well with a known estimate from strong-coupling expansions of the mass gap
$\tilde g \approx 1.12$ \cite{hamer_1979} and less well with an estimate $\tilde g \approx 1.05$ based on
Padé approximations of $\beta$-functions from strong-coupling expansions \cite{hornby_1985}, 
as well as a number of other methods \cite{allton_1988} that indicate $\tilde g \approx 1.00$ 
(the parameter given in these studies is usually $x = 2/\tilde g^2$).
It is worth noting that estimating the transition point of a BKT transition is notoriously
difficult due to the exponential scaling of the mass gap near the transition \cite{kosterlitz_1973}
and it is possible that our estimate would shift given data at larger bond dimensions, or by the use of
more reliable indicators than the onset of nonphysical symmetry-breaking.
An accurate determination of $\tilde g_{\text{BKT}}$ is, however, beyond the scope of this work.

Despite the impossibility of representing the ground state precisely
in the gapless phase at weak coupling, 
the scaling of von Neumann entropy and correlation length
in MPS ground states with a range of finite $D$ can be
used to estimate the central charge $c$ of the conformal field theory (CFT)
describing the phase \cite{tagliacozzo_2008, stojevic_2015}. We fit data for
$D=22,28,34,\dots,80$ at $1 / (\tilde g\sqrt{2}) = 0.75,0.8,0.85,0.9$ and find 
$c=0.992\pm 0.009$, matching the known result of $c=1$ for the 2D classical
XY model \cite{francesco_2012}, which is identical with the classical $O(2)$ rotor.

We now turn to the $O(4)$ rotor, which is known to exist in a single, gapped phase 
down to the weak-coupling limit $\tilde g \rightarrow 0$ \cite{sachdev_2011}.
Here, we expect our choice of basis to become increasingly bad as we approach
weak-coupling, due to the occupation of higher Fourier modes. We also expect
greater entanglement in the exact ground state as the potential term, coupling 
nearest-neighbor edges, begins to dominate, and the lattice correlation length grows. 
This is not enough, however, to explain
the very sudden occurrence of nonphysical symmetry-breaking. This is likely due
to the ``crossover'' phenomenon, a property of the $O(N > 2)$ models and of
nonabelian gauge theories,
referring to persistence of strong-coupling behavior up to a certain region of 
parameter space, where weak-coupling behavior rapidly takes over. 
Despite the sudden onset of the weak-coupling regime,
we still expect the nonphysical symmetry-breaking transition to disappear as 
$D \rightarrow \infty$, as we indeed observe in Figure \ref{fig:SB_extrap}.

\subsection{Mass gap}

\begin{figure}[t]
  \includegraphics[width=\linewidth]{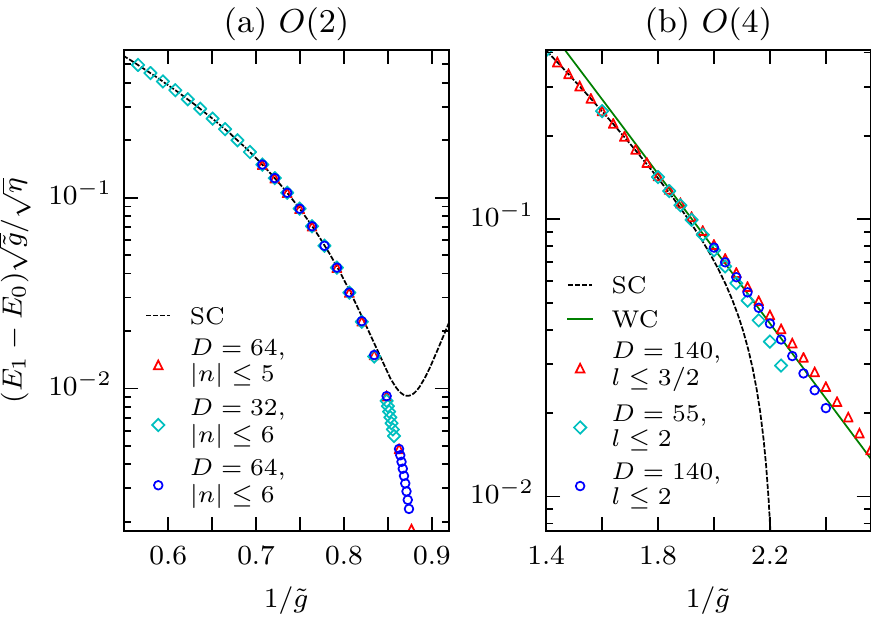}
  \caption{\label{fig:gaps} The MPS mass gap for (a) the $O(2)$ rotor
           and (b) the $O(4)$ rotor
           for bond dimensions $D$ and Fourier cutoffs in $|n|$ and $l$ respectively.
           The strong coupling expansion of \cite{hamer_1979} is shown (SC), as is the
           weak coupling result (WC) for the $O(4)$ case, which is known exactly \cite{hasenfratz_1990}. 
           In (b), the curves are adjusted by an anisotropy parameter $\sqrt{\eta}$ 
           to account for the renormalization of the speed of light \cite{shigemitsu_1981}
           (for $O(2)$, $\eta$ is set to one).
           Near the phase transition for $O(2)$, and as we enter 
           the weak coupling regime for $O(4)$, finite entanglement effects and,
           for $O(4)$, Fourier cutoff effects become important.
           }
\end{figure}

Our next source of information is the mass gap. Here, we can directly compare
our results with the results of 8th-order and 6th-order 
strong-coupling series expansions (SCE) for the $O(2)$ and $O(4)$ models, respectively \cite{hamer_1979}.
We find excellent agreement for both models up to the vicinity of the 
$O(2)$ phase transition and the $O(4)$ crossover region. Moving closer,
Figure \ref{fig:gaps} shows that the mass gap descends toward zero at a finite
coupling for $O(2)$, whereas for $O(4)$ the log-linear plot shows linear behavior,
indicating a finite mass gap for all finite couplings. For comparison, we plot
the exact asymptotic weak-coupling scaling for $O(4)$ \cite{hasenfratz_1990},
taking into account speed-of-light renormalization effects due to 
the stark space-time asymmetry of the Hamiltonian discretization \cite{shigemitsu_1981}.
We find very good agreement with the weak-coupling prediction, showing that we
are successfully entering the asymptotic scaling regime, although we also see from the
plot that finite entanglement effects start to limit the accuracy (the $D=140$ curve
remains accurate for longer than the $D=91$ curve for $l_{\max} = 2$), as indeed does the Fourier mode 
truncation (the $l_{\max} = 2$ curve is more accurate than the $l_{\max} = 3/2$ curve for $D=140$). 

In the $O(4)$ case it is also interesting to note that, holding the Fourier cutoff at
$l \le 2$, increasing the bond dimension appears to interpolate
between the SCE result and the weak-coupling
result. The effect is even clearer when more values of $D$ are
considered. This makes sense if we recall that the 
SCE for the mass gap perturbs the $\tilde g \rightarrow \infty$ ground state
and first excited state, both product states without entanglement in the
Fourier basis, by
repeatedly applying the nearest-neighbor term in the Hamiltonian up
to some order \cite{kogut_1975, hamer_1979}. The higher the order, the less local
the correlations introduced into the states will be. In the same way,
raising the bond dimension of an MPS ground state utilizing the
Fourier basis allows longer-range correlations to be represented. In
this sense, our MPS methods and SCE's are very
similar techniques, and it is not surprising that the MPS results at smaller
bond dimensions match low-order SCE results well.

Using similar reasoning, we can understand why the SCE results and the
lower-$D$ MPS results \emph{underestimate} the mass gap in the $O(4)$
weak-coupling regime. Given that the ground state in this region 
consists of highly non-local structures (Wilson loop
excitations of various sizes) \cite{kogut_1975}, limiting the order of
the SCE or restricting the amount of entanglement in the MPS should
both work against achieving these low-energy configurations, resulting
in an overestimation of the ground state energy. Indeed, we observe
significant differences on the order of $10^{-2}$ in the ground state
energy with $D$ as we enter the crossover regime.  Assuming the first
excitation is represented relatively accurately, this explains the
underestimation of the gap.

We note here that using a symmetric tensor
network ansatz \cite{mcculloch_2002, *singh_2010, *weichselbaum_2012}
might significantly extend the range of accessible effective bond
dimensions and so enable further penetration into the $O(4)$ weak-coupling
regime, although it would not allow access to the lowest-lying
excitations of the $O(4)$ model, which break the $O(4)$ symmetry.
For a model with truly local gauge symmetry, methods such as
that of \cite{buyens_2014} are required.

\begin{figure}[t]
  \includegraphics[width=1\linewidth]{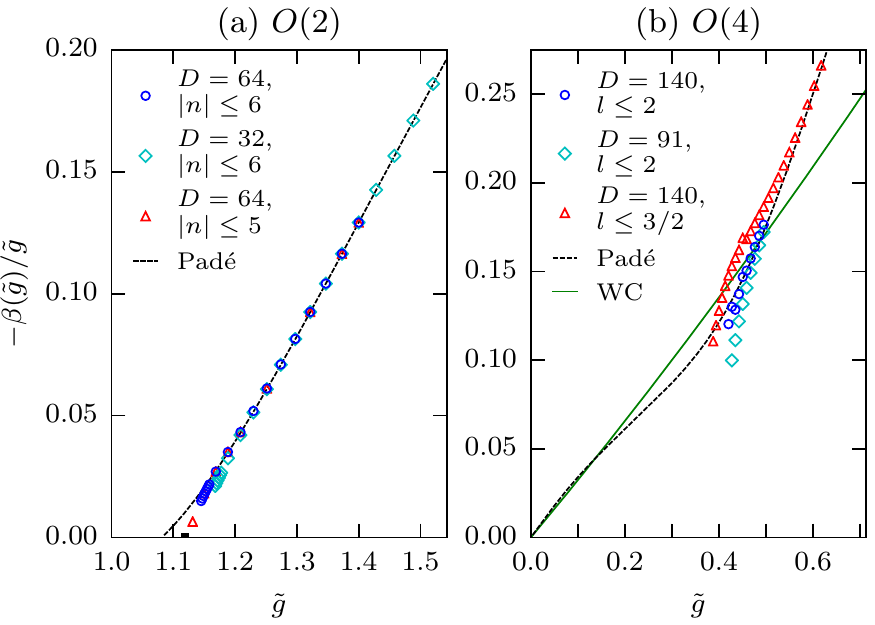} 
  \caption{\label{fig:betas} Beta functions determined from the MPS mass gap for (a)
           the $O(2)$ rotor and (b) the  $O(4)$ rotor, 
           together with a Padé approximant based on a strong-coupling expansion and the
           weak coupling result (WC) for the nonabelian case \cite{hamer_1979}.
           The estimate $\tilde g_{\text{BKT}}$ from Figure~\ref{fig:SB_extrap} of the 
           location of the $O(2)$ phase transition is marked in (a)
           using a black rectangle.
           For $O(4)$ it is clear from the
           $D=140$, $l_{\max} = 3/2$ curve that the numerical results begin to qualitatively
           follow the weak coupling behavior. However, there are clearly systematic
           errors present. This is expected because the beta function involves
           the numerical derivative of the mass gap, making it sensitive to
           small inaccuracies due to finite entanglement and Fourier mode truncation.
           }
\end{figure}

\subsection{Beta functions}

Using the mass gap and its first derivative in the coupling, 
one can calculate (see, for example \cite{hamer_1979}) 
the $\beta$ function as
\begin{equation} \label{eq:beta}
  -\beta(\tilde g)/ \tilde g = \left(1 - \frac{4}{\tilde g^2} \frac{F'(\tilde g)}{F(\tilde g)}\right)^{-1},
\end{equation}
where $F(\tilde g) = 2a(E_1(\tilde g) - E_0(\tilde g))/\tilde g$. 
Using finite-differences to compute $F'$, we may compute $\beta$ functions
from our mass gap results.

It is also possible to use an SCE for the mass gap to construct a Padé
approximant for the $\beta$ function. Furthermore, the $O(N > 2)$
weak-coupling behavior of $\beta(\tilde g)$ is known from perturbation
theory to be
\begin{equation}
  -\beta(\tilde g) = (N - 2) \frac{\tilde g^2}{2\pi} + (N - 2) \frac{\tilde g^3}{4 \pi^2},
\end{equation}
allowing this information to be incorporated, resulting in an
approximate $\beta$ function for all couplings for the $O(N > 2)$
rotor~\cite{hamer_1979}. 

We compare the Padé approximants
of~\cite{hamer_1979} with our MPS results in
Figure~\ref{fig:betas}, observing excellent agreement at stronger
couplings, with the numerical results deviating from the approximate
curve as we near the phase transition. In the case of $O(2)$, the
numerical data appears to predict a higher value for the phase
transition location than the Padé approximant, in good agreement with
our result from Figure \ref{fig:SB_extrap}. However, we also 
observe a shift in the results as the bond-dimension changes, with
the higher-$D$ results corresponding to a smaller prediction for 
$\tilde g_{\text{BKT}}$. This supports the possibility mentioned in
the previous section that using higher
bond-dimensions would result in a better correspondence with the
majority of literature results.

The $O(4)$ data ceases to
follow the Padé curve as we enter the crossover region, but does not
succeed in following the weak-coupling result accurately either.  This
is not unexpected, as both approximations are likely inaccurate in the
crossover region. We do, however, see large variations with $D$ and
$l_{\max}$, particularly as we near the nonphysical symmetry-breaking
transition. That errors are more visible for the $\beta$ function than
for the mass gap is expected since the numerical derivative amplifies
small errors in the mass gap.  We would need to reach higher bond
dimensions and Fourier mode cutoffs to achieve accurate results
further into the weak-coupling regime.  A further way of reducing
noise would be to compute the derivative $F'(x)$ analytically from the
MPS excited state.

\section{Concluding remarks}

We have shown that tensor network state (TNS) methods, in this case
uniform matrix product states (MPS), can successfully represent states of the
nonabelian quantum rotor model into the weak-coupling regime. The finite local basis,
achieved through Fourier-mode truncation successfully and efficiently 
captures strong-coupling physics, but becomes a more severe limitation at
weak couplings where, additionally, the spatial entanglement grows substantially.

This is promising for TNS approaches to pure nonabelian gauge theory, which is believed
to possess a very similar phase diagram to the $O(N)$ rotor models and,
on the ``hawaiian earring'' graph, is indeed equivalent to the rotor models studied here.
Our study also shows that high spatial entanglement is a feature of the 
theory from the crossover region onward, into weak-coupling. This may pose
a challenge for numerical approaches if it carries over to higher dimensional 
nonabelian lattice gauge theory, since large bond dimensions may be needed to access
the asymptotic scaling regime.

The author would like to thank Tobias J. Osborne, Leander Fiedler, Courtney Brell,
Karel Van Acoleyen and Kais Abdelkhalek for inspiring discussions. This work
was supported by the ERC Grants QFTCMPS and SIQS and by the cluster
of excellence EXC 201 Quantum Engineering and Space-Time Research.

%

\end{document}